# Declarative Modeling for Building a Cloud Federation and Cloud Applications

Giuseppe Attardi, Alex Barchiesi, Alberto Colla, Fulvio Galeazzi, Giovanni Marzulli, Mario Reale

Consortium GARR, via dei Tizii 6, Roma, Italy
{name}.{surname}@garr.it

**Abstract.** The paper illustrates how we built a federated cloud computing platform dedicated to the Italian research community. Building a cloud platform is a daunting task, that requires coordinating the deployment of many services, inter-related and dependent on each other. Provisioning, servicing and maintaining the platform must be automated. For our deployment, we chose a declarative modeling tool, that allows describing the parts that compose the system and their relations of supplier/consumer of specific interfaces. The tool arranges the steps to bring the deployment to convergence by transforming the state of the system until it reaches a configuration that satisfies all constraints.

We chose a declarative service modeling approach for orchestrating both the deployment of the platform by the administrators and the deployment of applications by users.

The cloud platform has been designed so that it can be managed by this kind of automation, facilitating the deployment of federated regions by anyone wishing to join and to contribute resources to the federation. Federated resources are integrated into a single cloud platform available to any user of the federation. The federation can also seamlessly include public clouds.

We describe the architectural choices, how we adapted the OpenStack basic facilities to the needs of a federation of multiple independent organizations, how we control resource allocation according to committed plans and correspondingly how we handle accounting and billing of resource usage.

Besides providing traditional IaaS services, the cloud supports self-service deployment of cloud applications. The cloud thus addresses the long tail of science, allowing researchers of any discipline, without expertise in system or cloud administration, to deploy applications readily available for their perusal.

## 1    Introduction

Scientists worldwide are increasingly adopting cloud technologies for carrying out computational tasks that require dealing with growing demands of compute and storage resources needed to cope with Big Data challenges. Cloud computing promises to help relieving scientists from the more mundane tasks of procuring, installing and provisioning the computing resources, of servicing and maintaining the basic system software up to date. Resources can be allocated on demand, scaling according to varying usage patterns and eventually reducing the cost for individual groups to maintain their own infrastructure.

Large research infrastructures are being planned at the EU scale, in the context of the ESFRI roadmap [27], aiming at providing scientists with suitable tools for scientific investigations. This requires fulfilling increasing demands on data volumes and computing power.

Projects and initiatives such as Indigo-Datacloud [14], EGI [17], European Open Science Cloud [15], HelixNebula [16] are addressing the implementation of cloud services for the European academic community.

Indigo-DataCloud is developing middleware to implement various cloud services ranging from authentication, workload and data management and is collecting a catalogue of cloud services. The project has just released its second software release, ElectricIndigo. The Indigo initiative puts its emphasis mainly on filling the gap between cloud developers and the services offered by existing cloud providers, rather than provisioning its own cloud service.

EGI coordinates a federated cloud, which originally relied on OCCI [18] and CDMI [29] as web services interfaces for accessing resources from OpenNebula [28] and OpenStack [1] clusters or from  public providers. The approach consists in providing an extra abstraction layer over resources provided by national grid initiatives which remain separate and independent of each other.

HelixNebula [16] is exploring how best to exploit commercial cloud providers in the procurement of cloud infrastructures for research and education. The approach consists in creating a private-public partnership for the procurement of hybrid clouds, and has recently started its third prototype phase, involving three contracting consortia.

The European Commission is promoting the European Open Science Cloud [15] as a general framework to support open science and research, covering many issues ranging from technical, to accessibility and governance issues, except for building a cloud infrastructure.

Many of these research funded projects either address the needs of specific communities, e.g. high energy physics, or provide services at the prototype or experimental level to a limited group of users, with a limited amount of resources, like those grouped within the EGI Federated Cloud initiative.

Moving from the prototype stage to a production stage, offering large amounts of resources to a large community is a challenge in terms of effort and resources. Building a well-supported and equipped cloud computing platform requires significant investments coming either from large commercial cloud providers or from public organizations that decide to invest in the far-reaching goal of setting up a real cloud infrastructure for science. One possible alternative to a central large funding approach is the federated approach, where the infrastructure is built bottom-up, by combining medium/large facilities into larger ones, to achieve an adequate scale.

Two initiatives that pursue this direction are the bwCloud project [19] by the state of Baden-Württemberg in Germany and the GARR Cloud project in Italy. They share the aim of providing a production quality cloud compute service targeted at 100,000+ potential users.

We designed our cloud computing platform based on the open source OpenStack platform, to meets the needs of massive automation, orchestration of resources, light deployment, involving a variety of underlying resources (Virtual Machines, LXD containers, bare metal servers), and deployment of cloud services in remotely managed data centers.

The community of our potential cloud users includes both large institutions with significant expertise and investment in the use of cloud technologies and small organizations with little or no hardware resources nor skills to undertake their own cloud deployments. The former are candidates for the use of Virtual Datacenters, built upon hardware resources that they may themselves supply. The latter will benefit from the availability of ready-made cloud applications that they can deploy with minimal effort.

The GARR Cloud offers three categories of services to address these different needs:

1. traditional IaaS through Virtual Machines in several flavors;
2. Virtual Data Centers consisting in a set of resources that an administrator can group into projects and assign to users;
3. Self-provisioning of PaaS cloud applications.

The third category aims to widen the audience of cloud computing services to researchers of any disciplines, addressing the needs of the so called "Long Tail of Science", enabling individuals or small research teams from any discipline to deploy self-service cloud applications, without requiring a deep expertise in cloud technologies or system administration. These applications are available from a public catalogue that includes already over 200 applications in several domains (e-learning, CMS, Big Data analysis, web development, LAMP, databases, etc.). These applications can be deployed by anyone with a few clicks through a web GUI.

The cloud federation builds upon resources pledged by partner institutions, and relies on a common reference architecture model for service provisioning.

The federated architecture can also include clouds from public cloud providers, allowing seamless integration of private and public clouds into a hybrid cloud to be able to respond, for instance, to temporary bursts of demand of cloud resources.

This article reports the relevant features of the design and deployment of the GARR Federated Cloud, involving authentication, workload management, storage provisioning, monitoring and operations.

## 2 Automation

We placed special emphasis on automation procedures for the installation, service configuration and management of the physical and virtual resources underlying a cloud platform.

After an assessment of several currently available Open Source automation tools, including Chef [33], Puppet [34], Ansible [], Heat [35], we decided to concentrate on higher level orchestration tools, in particular Juju and Brooklyn, which fit our preference for a declarative modeling tool.

### 2.1 Juju

Juju by Canonical [3] allows deploying complex applications to several cloud platforms as well as bare metal servers. Juju uses "charms", namely pre-packaged recipes defining requirements and steps involved at various stages of a service lifetime, from configuration, to installation, to upgrade. Charms also define configuration parameters and the interfaces they require or provide. The charm hooks to be execute at each step can be implemented in any language as well as through configuration management tools, such as Ansible, Chef, Puppet. A charm bundle is a collection of charms, that describes in a YAML file their properties and relationships. Juju contains command line client and web client.

### 2.2 Apache Brooklyn

Apache Brooklyn [26] allows modeling, monitoring and managing cloud applications with policy-based automation. For modeling, Apache Brooklyn uses a concept of *blueprint*, which defines an application in terms of components, their configurations, their relationships and deployment scenarios. A blueprint is written in YAML. Apache Brooklyn supports many applications out of the box and has web UI to deploy, monitor and manage applications. Also, Apache Brooklyn supports many deployment locations including cloud platforms or the existing nodes. The same blueprint can be used to deploy an application to different locations. Policies define unattended changes to an application, based on readings from application sensors.

We eventually settled on Juju, because, in combination with MAAS [4] for bare metal provisioning, it offers a single tool for managing the provisioning of both the cloud itself and of cloud applications, as summarized in this picture:

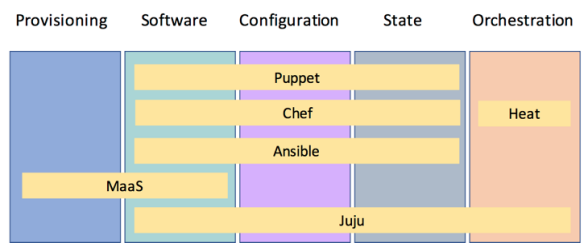

Juju indeed provides charms for all services required for installing the cloud platform as well as a large repository of bundles for deploying end user cloud applications.

Leveraging MAAS as a provider of bare metal servers, and Juju as an orchestrator for services over physical hosts, LXD containers and VMs, we can spawn both OpenStack infrastructural services at scale and cloud applications.

## 3    Declarative Service Modeling

We wished for our cloud platform to exhibit these features:

1. Declarative specification of services
2. Composability of services
3. Maximum agility
4. Abstraction from infrastructure and services
5. Deployment over multiple providers

An *imperative* service deployment model involves specifying *how* to perform the deployment providing an explicit *management plan* that explicitly describes each step to be performed. A *declarative* service deployment model involves specifying the intent of the service architecture, providing a high-level description of the requirements and constraints; a deployment engine interprets the description and takes care of generatimg and executing the plan to achieve the required configuration.

We chose a declarative service modeling approach for describing and deploying both cloud services and cloud applications. Services are described by means of a declarative domain specific language and deployed by a tool that interprets those specifications and produces and maintains a plan of actions to perform deployment. This is an alternative to deployment performed by procedural scripts, which specify how to do certain steps, but not why and what should be done. Maintaining those scripts requires overall knowledge of how all parts of the system behave, with high risks of missing important constraints which are nowhere expressed except in the heads of those who wrote them.

A declarative approach to service modeling allows specifying the logical composition of services into applications, and their horizontal scaling. For example, for an application $A$ that requires services $B$ and $C$, one can specify their requirements and inter-relations, e.g. that application $A$ requires $x$ GB of memory, $y$ vCPUs and $z$ GB of storage; that it needs to use the interface $I1$ provided by $B$ and interface $I2$ provided by $C$ and finally that it should be co-located with $B$.

The composability of services allows for users to share models, reusing components developed by others and building just the missing parts.

Agility is achieved by being able to quickly modify the configuration of a system and letting the deployment engine to determine a course of action to perform the necessary steps to achieve the new configuration.

The abstraction from underlying infrastructure and the ability to deploy on different providers is delegated to the tool itself that maps deployment actions into calls to the APIs of the providers, rather than requiring applications to be built using a special abstraction layer.

Endres et al. [22] compare two modeling patterns for automated deployment of applications in the cloud: declarative vs. imperative. They examine in particular the most utilized and established deployment technologies in modern cloud environments: the Cloud standard TOSCA[12], the technologies IBM Bluemix [32], Chef [33], Juju [3], and OpenTOSCA [31].

TOSCA enables modeling declaratively the application structure through Topology Templates, out of which the imperative provisioning logic is generated.

TOSCA however only deals with the portability of service Templates, portability of the ingredients of an IT service (especially the code artifacts) is not addressed by TOSCA. Similarly, mobility of data used by a corresponding service instance is not in the scope of TOSCA. TOSCA enables explicitly imperative provisioning: workflow models can be attached to services that implement the provisioning imperatively.

The academic open-source prototype OpenTOSCA [31] implements the TOSCA standard and contains a generator for BPEL workflows that allows to generate provisioning plans that can be customized individually for certain needs.

In the GARR Federated Cloud architecture, we use the declarative modelling of Juju bundles in order to provision both the federated cloud platform itself and the cloud applications that users can deploy in self-service.

### 3.1 Juju Deployment Model

Juju deployment model relies on a coordination engine and agents on each deployed node.

Juju *charms* define the events lifecycle an application goes through and the tasks to be executed corresponding to those events. The basic set of events include: `install`, `leader-elected`, `config-changed`, `start`, `update-status`. There are further events for each relation [name] ([name]`-relation-joined`, [name]`-relation-changed`, etc.) and for each storage pool ([pool]`-storage-attached`, [pool]`-storage-detaching`). Developers can also define their own states to which the charm can subscribe for triggering an appropriate action.

It is required that individual hook handlers are idempotent, so that running them more than once does not cause inconsistencies or divergence. Properly written charms ensure that a deployment that was stopped or blocked for any reason at any time, can be resumed safely to reach completion.

Charms also express required facilities and exposed interfaces by the charm. Bundles express constrains, configuration parameters and relationships between charms that provide/implement an interface.

The foundation for a declarative modeling of application deployment lays on the constraints and requirements expressed in bundles or charms and on the hooks defined in charms that specify proper reactions to certain changes in conditions of the system, notified through events. The Juju engine follows a reactive pattern, triggered by events that cause corresponding hook handlers to run. Multiple handlers may match for a given hook and will be run in a non-determined order. Running the handlers or issuing Juju commands may cause additional events. The state engine is evaluated every time an event occurs. The engine runs until convergence to a stable state.

### 3.2 Cloud Application Deployment

The GARR Cloud platform exploits Juju [3] also for allowing users to install self-service cloud applications. Juju charms (deployment recipes of a single service) or bundles (aggregations of services) can be selected from an online repository (jujucharms.com). User can compose applications by combining and connecting charms. New charms can be developed and contributed to the repository. We are actively contributing charms, in particular for use in education, like Moodle and Jupyter Notebook, and encourage our users to do so, sharing their solutions with others.

For example, we developed a Juju bundle for provisioning a PaaS service for Moodle, a popular e-learning platform. The application requires two components: Moodle itself, which in turns requires an Apache web server, and a PostgreSQL database backend. Their relationships are expressed through a declarative notation, either graphical or through a Yaml file. For deploying the application, the user sets a few configuration parameters and then sends the bundle to the Juju controller node, which interacts with the OpenStack API to perform the deployment. OpenStack carries out the tasks requested by the controller and deploys the requested VMs. The controller, once the VMs are available, installs agents on them with which it interacts to complete the installation and to react to any subsequent change of state. Thereafter, OpenStack simply provides the virtual infrastructure without being aware of the application roles assigned to each node.

Here for example is how we specify the constraints in the Juju bundle for Moodle that we developed:

```
applications:
 moodle:
    charm: "cs:~csd-garr/moodle"
    num_units: 1
    to:
      - 0
 postgresql:
    charm: "cs:postgresql"
    num_units: 1
    to:
      - lxd:0
    options:
      extra_pg_auth: host moodle juju_moodle 10.0.0.1/24 md5
relations:
 - ["postgresql:db", "moodle:database"]
machines:
 "0":
    series: xenial
    constraints: "arch=amd64 cpu-cores=1 mem=2048 root-disk=20480"
```

The Moodle application requires a PostgreSQL database. The recipes for their deployment can be obtained from the Charm Store (`cs:~csd-garr/moodle` and `cs:postgresql` respectively). A single machine is used for both (numbered "0" in the "machines" section), which should run the `xenial` version of the Ubuntu operating system and have the given hardware features. Moodle is installed on the physical machine (`--to 0`) while PostgreSQL is installed on an LXC container (`--to lxd:0`) on the same machine and authorization is granted to accept connections from the Moodle server. All details of the database configuration are hidden by the single "`relations`" line, which triggers the installation of database and the configuration of Moodle to use it.

Scaling the application just requires to change the number of units and adding an HA proxy for load balancing:

```
haproxy:
    charm: "cs:haproxy"
    num_units: 1
    expose: true
    to:
        - "4"
relations:
    - ["postgresql:db", "moodle:database"]
    - ["haproxy:reverseproxy", "moodle:website"]
```

The bundle can also be deployed interactively via the Juju GUI, configured and scaled to many cloud providers:

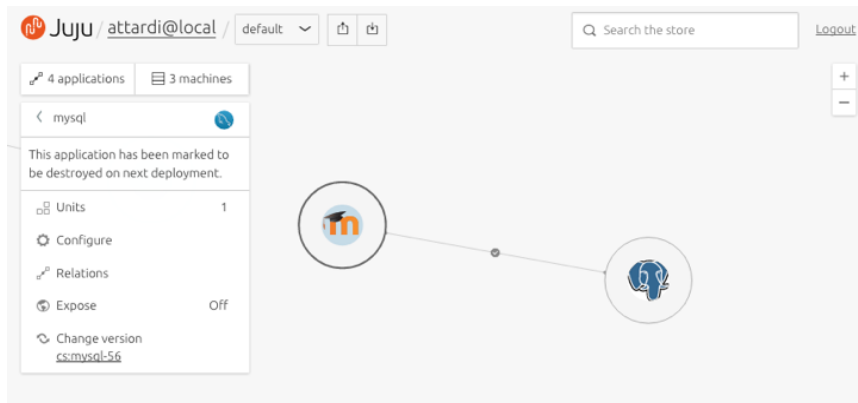

The federation can include also public cloud providers: the deployment mechanism is identical for both the private and public clouds. This avoids the risks of lock-in, since users can deploy their applications anywhere using a tool that is independent of the cloud provider.

Notice that the deployment mechanism is different from the traditional mechanism of exploiting pre-built VM images. In the latter case, the images are black boxes, without any visible information of what they contain and how they were built: therefore, it is difficult or impossible to perform upgrades or maintenance on these images. Instead our deployment through declarative bundles, provides a full description of how the bundle is created, therefore the application can be managed by the owner, who can take care of upgrades to the software packages used in the application.

### 3.3 Difference with Heat Templates

HEAT orchestration templates provide the means to automate the process of creating resources via small, human-editable scripts. They can deal declaratively with orchestration of OpenStack facilities, while the services to be provisioned on those facilities must be dealt imperatively by supplying installation scripts. In our previous example,

Heat templates could also help us automating the deployment of the two servers needed for the WordPress application [adapt to Moodle]. The services themselves would still have to be installed and configured manually.

```
heat_template_version: 2013-05-23
description: >
  This WordPress template installs two instances: one running a
  WordPress deployment and the other a MySQL database
parameters:
  key_name:
    type: string
    description: Name of a KeyPair to enable SSH access
    default: test_key
  instance_type:
    type: string
    description: Instance type for web and DB servers
    default: m1.small
  image_id:
    type: string
    description: Image to use for the WordPress server.
    default: fedora-20.x86_64
  db_name:
    type: string
    description: WordPress database name
    default: wordpress
  db_username:
    type: string
    description: The WordPress database admin account username
    default: admin
    hidden: true
  db_password:
    type: string
    description: The WordPress database admin account password
    default: admin
    hidden: true
resources:
  DatabaseServer:
    type: OS::Nova::Server
    properties:
      image: { get_param: image_id }
      flavor: { get_param: instance_type }
      key_name: { get_param: key_name }
      user_data:
        str_replace:
          template: |
            #!/bin/bash -v
            yum -y install mariadb mariadb-server
            touch /var/log/mariadb/mariadb.log
            chown mysql.mysql /var/log/mariadb/mariadb.log
            systemctl start mariadb.service
            # Setup MySQL root password and create a user
            mysqladmin -u root password db_rootpassword
            cat << EOF | mysql -u root --password=db_rootpass-
word
```

```
                CREATE DATABASE db_name;
                GRANT ALL PRIVILEGES ON db_name.* TO "db_user"@"%"
                IDENTIFIED BY "db_password";
                FLUSH PRIVILEGES;
                EXIT
                EOF
            params:
                db_rootpassword: { get_param: db_root_password }
                db_name: { get_param: db_name }
                db_user: { get_param: db_username }
                db_password: { get_param: db_password }

  WebServer:
    type: OS::Nova::Server
    properties:
      image: { get_param: image_id }
      flavor: { get_param: instance_type }
      key_name: { get_param: key_name }
      user_data:
        str_replace:
          template: |
            #!/bin/bash -v
            yum -y install httpd wordpress
            sed -i "/Deny from All/d" /etc/httpd/conf.d/word-
press.conf
            sed -i "s/Require local/Require all granted/"
/etc/httpd/conf.d/wordpress.conf
            sed -i s/database_name_here/db_name/ /etc/word-
press/wp-config.php
            sed -i s/username_here/db_user/ /etc/wordpress/wp-
config.php
            sed -i s/password_here/db_password/ /etc/word-
press/wp-config.php
            sed -i s/localhost/db_ipaddr/ /etc/wordpress/wp-con-
fig.php
            setenforce 0 # Otherwise net traffic with DB is dis-
abled
            systemctl start httpd.service
          params:
            db_name: { get_param: db_name }
            db_user: { get_param: db_username }
            db_password: { get_param: db_password }
            db_ipaddr: { get_attr: [DatabaseServer, networks,
private, 0] }

outputs:
  WebsiteURL:
    description: URL for Wordpress wiki
    value:
      str_replace:
        template: http://host/wordpress
        params:
          host: { get_attr: [WebServer, networks, private, 0] }
```

In order to get the desired WordPress service configuration, the template specifies the requirements for the two Nova instances. The steps for installing and configuring the services to be run on those instances are included in templated code scripts in the `user_data` parameter of each server `properties` field.

The drawbacks of this approach are that: each installation of a service must include the code for its configuration, hence a change in a new release of the service would require modifying all the scripts that install the service; that services that use each other facilities are deployed independently, hence a change in one is not propagated to the other and even activation mail fail since deployment occurs in an unspecified order.

## 4    Cloud Platform Architecture

The GARR Cloud architecture aims at achieving these goals:

- Replicability and scalability;
- Integration of resources managed by different organizations;
- High level of automation in deployment and maintenance;
- Elastic allocation of resources assigned to a service, also exploiting resources shared by other regions;
- Both federated (SAML) and open (OIDC) authentication;
- High degree of reliability;
- Flexible security policies both at the resource and user levels.

### 4.1    Cloud Region Architecture

The reference architecture for a cloud region deploys OpenStack core services, in high availability through replication, and provides computing nodes, resilient storage (both block storage and object storage), and networking. The major components of the architecture are illustrated in Figure 4.

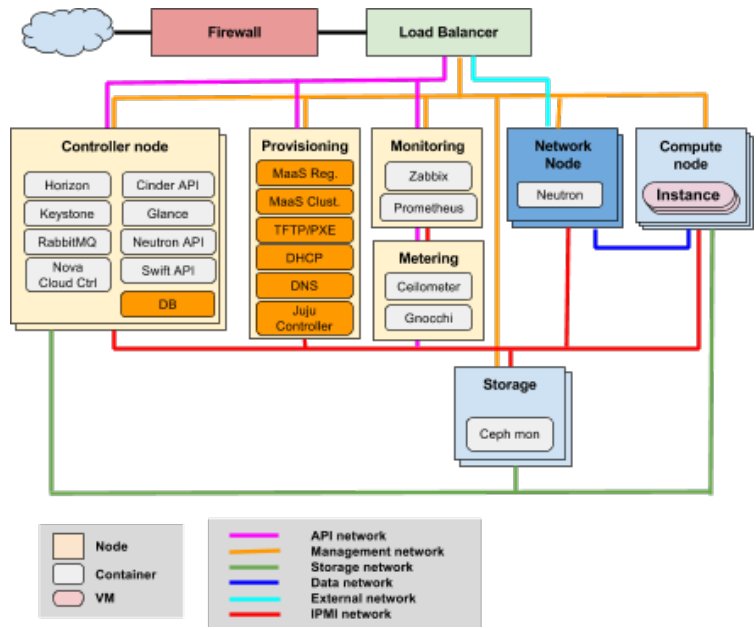

**Fig. 4.** Reference Architecture for OpenStack.

### 4.2 Federated Architecture

OpenStack does not have native facilities for building federations among different cloud service providers. A blueprint [13] currently exists for exploiting Availability Zones in a federated cloud, but the project has not yet started.

A more complex and more scalable solution is being proposed by Tricircle [12], which offers a OpenStack API gateway and networking automation to allow managing several instances of OpenStack as a single cloud, scattered over one or more sites or in a hybrid cloud.

Other approaches to federation are those based on the Open Cloud Computing Initiative (OCCI) [18], that introduces an interface layer that isolates the APIs of the various providers. OCCI defines a flexible API with a strong focus on integration, portability, interoperability and innovation while still offering a high degree of extensibility.

The Indigo-DataCloud project [14] is developing a data and computing platform targeted at scientific communities, deployable on multiple hardware and provisioned over hybrid, private or public, e-infrastructures. Indigo provides enhanced resource virtualization through TOSCA on Heat and OCCI support.

The project SeaClouds [10] aims at homogenizing the management over different providers and at supporting the sound and scalable orchestration of services across them. It uses two OASIS standards initiatives for cloud interoperability, namely CAMP [11] and TOSCA [12], which allow describing the topology of user applications independently of cloud providers, providing abstract plans, and discovering, deploying/reconfiguring, and monitoring applications independently of the peculiarities of the cloud providers. The solution relies on a deployer API that sits on top the Deployer Engine, which in turn drives the specific adapter for each cloud provider.

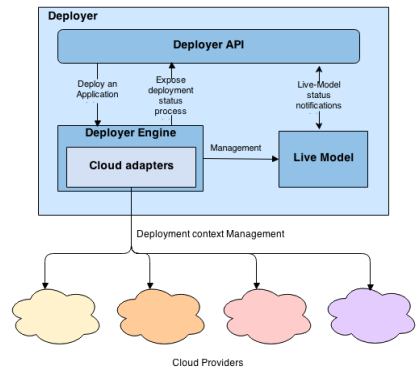

**Figure 6.** Architecture of the SeaClouds Deployer.

### 4.3 Proposed solution

The solution proposed in this paper is a very practical one, that avoids imposing an additional middleware layer and requirements on the way cloud applications are built and deployed. We assume that the federation is homogeneous, with all regions adopting OpenStack for deployment, which is a de facto standard. Access to services by other providers is achieved by mapping the OS API to those of the various providers. A dozen of mapping are available for the major public cloud providers, including AWS, Azure, Google, Rackspace and others.

Hence the federation is based on a multi-region OpenStack model, which is practically feasible assuming to federate clouds designed and implemented in a coordinated manner, based on compatible OpenStack platforms. In particular, there is a single Keystone service and a single dashboard for the whole federation.

The solution leverages the OpenStack mechanisms to scale to thousands of nodes and to expand onto different data centers and geographical areas, exploiting the concepts of Region and Availability Zone.

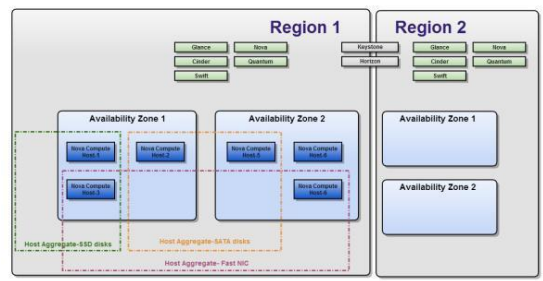

**Figure 2.** Example of Availability Zones in a multi-region architecture.

**Region.**

Each Region has its own deployment of OpenStack, including its endpoint APIs, network and computing resources, which are linked with other regions into shared centralized services such as identity and dashboard.

The concept of regions in OpenStack is flexible; we use it to group the resources that a partner decides to commit to the federation.

**Availability Zone.**

Within each Region, nodes are logically grouped into Availability Zones (AZ): when a VM is provisioned, one can specify in which AZ the instance should be started, or even within which specific node inside an AZ.

**Host Aggregate.**

Within a Region machines can be grouped into Host aggregates. We use host aggregates to distinguish hosts with different features, for example having different types of CPUs or SSD disks instead of HDD. The deployment of a VM can specify the requirement of a particular property, specified for example in the description of a given VM flavor.

## 4.4     Segregation and Sharing

The logic of the federation allows each organization to maintain control over the use of its resources, which part to keep for private use and which part to give for sharing with the federation. Technically this is done by associating to each region its own Keystone Domain, through which the local manager will administers its resources.

### 4.1.1 Domains and Hierarchical Projects.

Projects are the units of resource assignment and accounting. They can be arranged into a hierarchy, whose root is a domain. We use nested quotas, an obscure feature of some OpenStack services (e.g. Nova and Cinder), to control the usage of resources assigned to a project. Once a project hierarchy is created in Keystone, nested quotas let you define how much of a project quotas to assign to its subprojects. This feature is used to provide hierarchical multitenancy in the federation.

For example, in the situation shown in Figure 3, two separate domains are assigned to different organizations. Each organization is internally structured into projects and subprojects. Subprojects inherit role assignments, so for example the administrator for project Marketing is also an administrator for sub-projects National and International. Through nested quotas it is possible to control how the compute and storage resource can be divided among the two subprojects.

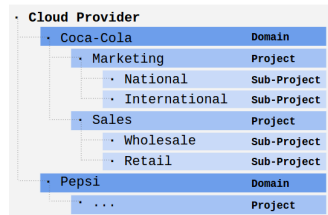

**Figure 3.** Hierarchical projects.

This model fits quite well also with the accounting mechanism of Amazon Organizations, which has been recently introduced for the purpose of accounting and billing the usage of resources.

### 4.5    Implementing the Federation through MAAS Regions

The high-level reference architecture of the federation involves multiple cloud regions, possibly managed by different research or educational organizations.

The architecture relies on OpenStack cloud technology, tailored to support the requirements of a federation among cooperating but independent organizations.

The architecture of the federation entails one Master Region, that hosts the common services of the cloud platform, and several Regions that host and provide additional resources to the federation.

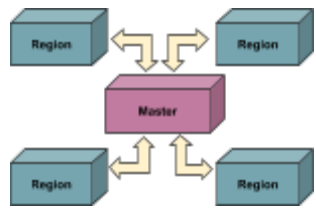

**Figure 1.** Architecture of the multi-region federated cloud.

The federation can seamlessly integrate also with public clouds for obtaining additional resources where to deploy cloud services.

The architecture design allows for scaling easily both within a region and globally. Adding capacity to a region is as simple as modifying the configuration of the bundle, by adding further compute nodes or storage, and then letting Juju take care of propagating the change to all affected services.

A new cloud region is added to the Master Cloud region only after successful deployment and validation of its OpenStack services. A candidate new region is first registered in a validation cluster, by inserting users and endpoints in a special Keystone database. After successful validation, endpoints are automatically inserted into the production Keystone database.

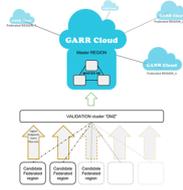

Authentication and authorization services are common to whole federation, implying users are globally recognized and therefore can be allowed to use any resource.

### 4.6    Central Authentication

Authentication and authorization in the federation are provided by a single central service that exposes a single API and a single web dashboard for the whole federated cloud. Since authentication itself is performed in a federated form, users are globally recognized and therefore they can be granted access to resources on the whole federation. For efficiency of access, the Keystone database is replicated in slave mode in each region.

This central model is simple and effective and is used both in the FIWIRE [5] federation and in the bwCloud project.

**Authentication and Authorization model.**
Access to the cloud platform is granted to users and administrators through three sources of authentication:

- — Local Keystone credentials for administrators
- — Federated SAML-based authentication delegated to the institutional identity providers of the user home organization (e.g. IDEM [2] or eduGAIN [20])
- — OIDC authentication (e.g. Google)

OpenStack provides hooks to implement SAML based authentication and authorization via the Keystone-SAML mapping mechanism, that maps remote identities to local Keystone users, thus allowing their assignment to groups and projects. Keystone itself is configured as a SAML Service Provider for the identity federation.

We choose to only delegate authentication to external federated authentication providers, while authorization is dealt by Keystone itself. Hence the registration procedure creates local users in a single default Keystone domain, and maps them to their federated identities, based on the `eduPersonPrincipalName` attribute (ePPN, from the eduPerson schema) released by the Identity Provider.

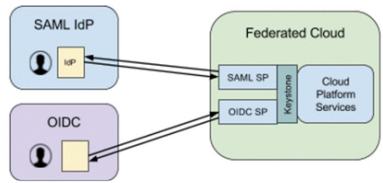

**Fig. 4.** Sketch of the federated authentication.

### 4.7    Storage

Each Region manages its own storage resources. For medium and large sized regions, we advise to use Ceph [21], a software-defined storage solution deployed at nearly 40% of production-level OpenStack installations, according to the OpenStack report 2016 [23].

Ceph can be installed and setup through Juju: the compute nodes are configured as disk servers (OSD) while the monitoring nodes (MON) are installed as LXD containers on the controller nodes. Ceph interfaces to Cinder for the provisioning of block devices, and to Swift for object storage.

Based on more than 2 years of operation, our experience with Ceph is overall positive in terms of robustness, data safety, aggregate performance and ease of maintenance.

Certain applications though, e.g. large scale analytics based on Hadoop Map/Reduce, with high storage performance demands, are not well served by Ceph. To meet these demands, we plan to introduce Host Aggregates with a pool of disks managed by a LVM Cinder backend, so that volumes can be attached directly to the compute nodes. Additionally, we plan to upgrade our storage by adding more solid-state disks.

### 4.8    Support in Deploying the Federation

We are committed to provide our partners with guidelines (`https://cloud.garr.it/doc/federation/`) and tools for automating the process of deploying cloud regions, joining the federation and managing the services of the cloud platform. Installing and maintaining a sizeable cloud may require significant manpower, unless automation is used.

We provide a recipe for the installation, deployment and maintenance of the cloud platform and the deployed services, starting from the hardware resources. Automation exploits the approach of services, which allows quick and easy deployment and management of services (whether it is a cloud infrastructure like OpenStack, or a workload such as Hadoop), connecting those services, and quickly scaling them up or down, without disruption to the running cloud environment.

Predefined software bundles and installation, configuration scripts are made available as open source, so that anyone can reproduce the process of creating a cloud from scratch and joining the federation.

## 5    Automation and Elasticity

Emphasis has been put on automation at all levels of the management and operation of the federated cloud infrastructure.

Bare metal servers are provisioned to the infrastructure and services installation and configuration tools by means of MAAS (Metal-as-a-Service) [4], which allows commissioning, decommissioning, tagging, network configuration including VLANs set up in an automated fashion, via IPMI, ensuring fast check-in and check-out of servers to the cloud infrastructure. This allows managing servers in an elastic way, in a similar way one would do with VMs. Additionally, MAAS can also manage Virtual Machines through their underlying hypervisor management channels.

Juju charms either already existed or we built ourselves for all the OpenStack services used in our architecture, allowing us to fully automate its deployment, with just one single command: `juju deploy bundle`.

Where to deploy each unit can be either left for Juju to decide automatically or it can be specified in the bundle, specifying whether to use a bare metal server, a VM or an LXD container: deployment can thus be tuned to meet optimal resource requirements.

Scaling a service horizontally requires just using the command `juju add-unit`, which triggers the allocation of new resources and to link them according to required relationships.

Application deployment through charms can be done via either the command line or a web GUI. The Juju GUI allows setting the configuration parameters of each service, connecting to other services according to the service provisioning requirements, and scaling each service to a suitable number of units.

Juju applications can be deployed on different cloud platforms, both private (OpenStack or local containers) and private, by installing a Juju controller on each cloud.

The logical service architecture of each cloud region is shown in Figure 1:

- Physical: consists of the physical hardware and network resources
- Operating System: corresponds to the services provided by the operating system
- Virtualization: provides an abstraction of OS services through virtualization
- Application Services: services provided by the platform

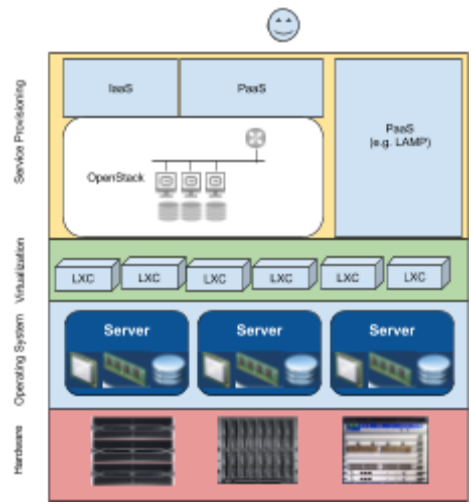

## 5.1 Juju for Cloud Infrastructure Deployment

We used Juju to deploy the full OpenStack infrastructure over bare metal servers. Juju offers the ability to deploy services into LinuX Containers (LXC). In the GARR Cloud all OpenStack modules and auxiliary components (database, HA proxy, etc.) are hosted on LXC placed on physical servers (while staying independent) for creating a highly reliable and load balanced infrastructure. For example, as the need arises containers can be moved from one server to another.

At this layer Juju interacts with MAAS which enlists the bare metal resources. MAAS itself is capable to handle geographically distributed resources thanks to its tiered architecture (a central region controller and distributed rack controllers): With Juju and MAAS it is therefore possible to distribute critical services in different data centers to achieve high level of resilience. All the core services of the GARR Cloud (database, keystone, etc.) are in HA on the GARR sites.

The services by the automation tools (MAAS controller, Juju bootstrap node, DHCP) are hosted on LXC containers as well.

## 5.2 Multi cloud support.

Juju operates through controllers which interact with a specific cloud provider, and models, which are operational environments to control a set of resources.

By deploying a Juju controller on a public cloud provider, Juju can carry out deployments via charms and bundles in that environment.

## 6    Current Deployment

The GARR Cloud is currently deployed in 5 datacenters in southern Italy (Napoli, Bari, Catania, Palermo, Cosenza), within GARR Network PoPs, interconnected at 100 Gbps over the GARR backbone. The infrastructure consists overall of about 9000 vCPUs, 10 PB of storage. Half of these resources are being dedicated to the cloud platform.

### 6.1 Early users feedback

The GARR Cloud has been operational since December 2016 and it is accessible from the landing page: garr.cloud.it. The service is offered to all members of the GARR user community with a six-month free trial period.

Over 350 users have registered for using the cloud in the last four months, and more are registering every day. Currently there are over 1000 vCPUs in use and over a dozen of Virtual Datacenters in operation.

A survey on early users offered the following insights:

— 80 % of users stated to be committed to continue using the service
— 60 % of users considered interesting the PaaS self-service deployment service
— 95 % of users rated the IaaS VM service as good or excellent
— 98 % of users rated the IaaS Virtual Datacenter as good or excellent.

A notable use of the GARR Cloud is by the InfraScience group at the NeMIS laboratory of ISTI CNR, which is heavily using the cloud (about 500 vCPU in total), in two projects:

• D4Science: integrated technologies that provide elastic access and usage of data and data-management capabilities, https://www.d4science.org
• OpenAIRE is the point of reference for open science in Europe, https://www.open-aire.eu

The GARR Cloud resources were instrumental in expanding the D4Science computing services, used by the scientific community worldwide to run both batch and interactive analysis tasks. Because these computations are resources hungry and can run for hours and even days, the underlying infrastructure must be fast and reliable.

OpenAIRE uses the cloud resources in many ways. Several slave build nodes are involved in the continuous integration system, some ElasticSearch clusters receive and handle data from several data gathering processes. A new Hadoop cluster is being built to migrate workflows to the cloud.

The group plans to further increase the use of GARR Cloud resources to improve the availability of its services and for mirroring our backup storage.

## 7 Related Work

The closest to ours, among federated cloud platforms for research, is the service deployed by the state of Baden-Württemberg and several local universities through the bwCloud project [19] which aims to create a federated Community Science Cloud. The architecture of the bwCloud relies on OpenStack and on the use of a single Keystone service for authorization, while the authentication is delegated to Eduroam-like service among the affiliated universities. The current deployment is at a smaller scale than ours, including 32 servers located in four regions.

The project is studying a suitable governance scheme for managing a federated system which aims to support 20,000+ users. The scheme must take into account the needs and interests of different stakeholders as well as of different funding agencies. Coordination within bwCloud therefore includes the development of a robust business model and procedures that enable user participation and satisfaction.

SeaClouds [10] aims at homogenizing the management over different providers and at supporting the sound and scalable orchestration of services across them. SeaClouds will inherently support the evolution of systems developed with it, coping with any needed change, even at runtime. The development, monitoring and reconfiguration via SeaClouds include a unified management service, where services can be deployed, replicated, and administered by means of standard harmonized APIs such as CAMP specification and Cloud4SOA project.

Apache Brooklyn [26] is an open source framework for modeling, monitoring and managing applications for cloud environments. Apache Brooklyn can manage provisioning and application deployment, monitor an application's health and metrics, understand the dependencies between components and apply complex policies to manage the application defined in a blueprint.

INDIGO-DataCloud is developing a data and computing platform targeted at scientific communities, deployable on multiple hardware and provisioned over either private or public infrastructures. By filling existing gaps in PaaS and SaaS levels, Indigo-DataCloud aims to help developers, resources providers, e-infrastructures and scientific communities to overcome current challenges in cloud computing, storage and networking.

Nectar Cloud [30] provides computing infrastructure, software and services that allow Australia's research community to store, access, and run data, remotely, rapidly and autonomously. Nectar Cloud's self-service structure allows users to access their own data at any time and collaborate with others from their desktop in a fast and efficient way. Nectar Cloud uses OpenStack to automate deployment including auto-scaling and load balancing and provides Virtual Machines from 1 to 16 Cores with 64GB of Ram.

## 8 Conclusions

The GARR Cloud platform offers computing resources to a large community of researches and we expect it to further grow by adding new regions to the federation by partners that are already planning to join.

The platform is being further developed in several aspects: billing and accounting are currently being designed and represent an essential step towards the full enrollment in production for the user community, moving beyond the early users and pilot experience.

There is potential opportunity to make the GARR Cloud as part of a larger European federation, leveraging the eduGAIN identity interfederation to authenticate users at a EU continental scale.

In later releases we plan to include also a Container Platform, based on Docker Engine or Kubernetes.

**Acknowledgments**
Members of the Cloud DevOps Team at GARR that participated to the development of the cloud service are: Alex Barchiesi, Alberto Colla, Fulvio Galeazzi, Gianni Marzulli, Mario Reale, Davide Vaghetti, Valeria Ardizzone, Paolo Velati. The members of the GARR Cloud Task Force participated to the design of the reference cloud architecture: Marco Aldinucci, Massimo Coppola, Paolo De Rosa, Giovanni Ponti, Davide Salomoni. Andrea Dell'Amico is a heavy user and tester of the platform and kindly described his experience.